\documentclass[conference]{IEEEtran}

\usepackage[usenames,dvipsnames,table]{xcolor}
\makeatletter
\def\ps@IEEEtitlepagestyle{%
  \def\@oddfoot{\mycopyrightnotice}%
  \def\@evenfoot{}%
}
\def\mycopyrightnotice{%
  {\footnotesize XXX-X-XXXX-XXXX-X/XX/\$XX.00~\copyright~20XX IEEE\hfill}
  \gdef\mycopyrightnotice{}
}

\usepackage{blindtext}
\usepackage{eso-pic}
\IEEEoverridecommandlockouts
\usepackage{cite}
\usepackage{amsmath,amssymb,amsfonts}
\usepackage{algorithmic}
\usepackage{graphicx}
\usepackage{textcomp}

\def\BibTeX{{\rm B\kern-.05em{\sc i\kern-.025em b}\kern-.08em
    T\kern-.1667em\lower.7ex\hbox{E}\kern-.125emX}}
    
\usepackage{eso-pic}
\newcommand\AtPageUpperMyright[1]{\AtPageUpperLeft{%
 \put(\LenToUnit{0.17\paperwidth},\LenToUnit{-2cm}){%
     \parbox{0.9\textwidth}{\raggedleft\fontsize{8}{11}\selectfont #1}}%
 }}%
\newcommand{\conf}[1]{%
\AddToShipoutPictureBG*{%
\AtPageUpperMyright{#1}
}
}    

\definecolor{gray20}{gray}{0.8}
\definecolor{gray5}{gray}{0.95}
\definecolor{olivegreen30}{RGB}{155,187,89}	

\definecolor{precolor_testpassed}{RGB}{118, 184, 42}
\colorlet{testpassed}{precolor_testpassed!38.1966}

\definecolor{precolor_testfailed}{RGB}{215, 0, 45}
\colorlet{testfailed}{precolor_testfailed!38.1966}

\usepackage{float}
\usepackage{subfigure}
\usepackage{soul}

\usepackage{orcidlink}
\usepackage{hyperref}

\begin{document}

\title{\vspace*{1cm} Appraisal of a Random Bit Generator Utilizing Smartphone Sensors as Entropy Source\\
}
\author{\IEEEauthorblockN{1\textsuperscript{st} Stefan Kutschera \orcidlink{0000-0002-7547-5137}}
\IEEEauthorblockA{\textit{Institute of Software Technology} \\
\textit{Graz University of Technology	}\\
Graz, Austria \\
stefan.kutschera@ist.tugraz.at}
\and
\IEEEauthorblockN{2\textsuperscript{nd} Wilhelm Zugaj}
\IEEEauthorblockA{\textit{Internet Technologies \& Applications} \\
\textit{FH JOANNEUM Gesellschaft mbH}\\
Kapfenberg, Austria \\
wilhelm.zugaj@fh-joanneum.at}
\and
\IEEEauthorblockN{3\textsuperscript{rd} Wolfgang Slany}
\IEEEauthorblockA{\textit{Institute of Software Technology} \\
\textit{Graz University of Technology}\\
Graz, Austria \\
wslany@ist.tugraz.at}
}

\maketitle
\conf{\textit{  Proc. of the International Conference on Electrical, Computer, Communications and Mechatronics Engineering  (ICECCME) \\ 
16-18 November 2022, Maldives}}


  \begin{abstract} 
We aim to access entropy sources available within smartphones in order to construct and evaluate a random number generator which is competitive in comparison with existing and proven random number generators.
A prototype utilizing the herein proposed algorithm shall generate data that can be tested against the Statistical Test Suit provided by NIST.
Although our initial intention of using cosmic radiation failed, we were able to extract randomness from incoming video and audio sources. We found that it is possible to access these sources of entropy utilizing sensors from smartphones resulting in 15 out of 15 successful passed tests within the Statistical Test Suit. 
We also found that wrong methods of sensor data collection using our prototype eventually generates weak random numbers and fails NIST's  Statistical Test Suit. Finally, we suggest that in order to reach the initial goal of providing a smartphone-based true non-deterministic random number generator the detection of muons shall be researched.
\end{abstract}


\section{Introduction}\label{chap:intro}
\textit{,,Die Quantenmechanik ist sehr achtunggebietend. Aber eine innere Stimme sagt mir, daß das noch nicht der wahre Jakob ist. Die Theorie liefert viel, aber dem Geheimnis des Alten bringt sie uns kaum näher. Jedenfalls bin ich überzeugt, daß der nicht würfelt.'' {Albert Einstein}} 


Random numbers are the bread and butter of many cryptographic primitives. They are used for example in cryptographic protocols, to generate keys, as seeds or to generate big prime numbers only to name a few. Thus, the easy availabilty of TRNGs is mission critical. Cryptography does not only need cryptographically strong algorithms, it also needs a strong source of randomness.

Albert Einstein's Special Relativity Theories and later the General Relativity Theory were since his proclamation proven many times. Interestingly, the muon proved time dilation and length contraction once again. The above shown untranslated German quote from Einstein was written 1926 in a letter to German Mathematician and Physicist, Max Born stating that 'he', thus God, does not play dice. Based on what we know about randomness today, it seems, Einstein was not right at all. For example, quantum mechanical properties within a single photon source are exploited in order to create a long sequence that has no repetition patterns and passes several statistical tests.\cite{idquantique:2016}

\subsection{Problem Statement}\label{sec:problem}
Utilizing quantum mechanics on a single photon source a quantum random number generator (QRNG) is the choice if a true random number generator (TRNG) is needed. As bizarre and futuristic as this technology may sounds it is freely available and can be bought from businesses as well as the general public. However, the ''Quantis USB Legacy`` QRNG from IDQuantique costs currently around 1.200 EUR and has a volume of 215$cm^3$, which might not fit in everyones pocket. Current smartphones occupying around 80,2 to 113$cm^3$ space and are already in pockets of manifold people. Inspired by those facts this work tried to construct a random number generator (RNG) based on everyday carry-on items, hence smartphones. Already available sensors on smartphones can be used to access a strong entropy source. With smartphones being available every other corner, our RNG is not only highly available but also inexpensive. Moreover, it could function as a basis to ensure secure cryptographic algorithms as used in key management within communication, transaction or encryption.  \cite[p.23]{80057rev5:2020}

\subsection{Research Questions}\label{sec:rq}
With the underlying work during the author's masters thesis research \cite{kutschera:2019}, the aim of this paper shall as well answer the following questions:
\begin{itemize}
    \item How can a random bit generator be constructed by using smartphone sensors as an entropy source based on existing approaches?
    \item How does the evaluation of the proposed algorithm compare to existing random bit generators?
\end{itemize}

\subsection{Hypothesis}\label{sec:hypothesis}
The usage of RNG's can be found massively in ones everyday internet life. An example is the Hypertext Transfer Protocol Secure (HTTPS), where the Transport Layer Security (TLS), and further the utilized Rivest Shamir Adleman (RSA), use new random numbers for every secure connection to a website. 

In order to address this problem of having a secure and cheap RNG at hand, this work will try to create a prototype of a widely available random bit generator (RBG) that can be used to create a strong random bit sequence, that shall be considered immune against statistical attacks when tested against NIST Statistical Test Suite.\par

We developed a proof of concept application that utilizes one or more sensors available on common smartphones. We defined the application as successful if the resulting random bit sequences was immune against statistical attacks. Due to its attributes, we called this application Economic Random Bit Generator (ERBG). 

\subsection{Methodology}\label{sec:method} 
It is well known that there are many other RNGs available. Some of them also use smartphone sensors as their entropy source. Thus, we did a literature survey of the most similar and relevant RNGs first. Based on that we developed a proof of concept that is not only limited to single picture frames or color channels but uses the entropy within a whole video file. Finally, the file output of our proposed ERBG was evaluated against the Statistical Test Suite from the National Institute of Standard and Technology in the United States of America (NIST) \cite{80022:2010}.

\section{Related Work}\label{chap:related}

\subsection{History \& Definition of Randomness}\label{SEC_random_historic} 
The term 'random' itself appeared in the Middle Ages and was attributed to supernatural forces. \cite{batanero:1998} 
During the same period, objects such as dice or astragali (i.e., bones used as an oracle) were used to predict the future and prevent any advantages for participants in games. The quote of the Greek philosopher Leucippus in the 5th Century B.C. ``Nothing happens at random; everything happens out of a reason and by necessity'', which provides an  insight into the ideologies during this epoch. \cite{bennet:1993}
On the other hand, distinguished random phenomena about the probability that a calculus can give some information and random phenomena for which there was no possibility of prediction until the law behind was fully discovered. \cite{poincare:1936}

The term 'random' can be divided into two groups, formal and informal. \cite{batanero:1998}
As an informal term 'random' is used to explain causes that are not fully explainable by known physical laws or by the general public for causes that are too complex at a first glance. The same term used in a formal matter, the definition states the sequence under analysis has maximum complexity or, in other words, a lack of patterns within the sequence. \cite{batanero:1998} 
Subsequently, it can be argued that a sequence is random when it cannot be predicted. Secondly, a random sequence contains a maximum of information. Thus, it cannot be compressed.

\subsection{Random Number Generators}\label{SEC_rng_types}
In cryptographic schemes, a randomly generated number is often required. The term RNG is often unanimously used with RBG since the binary output of RBGs is eventually transformed into other numerical systems thus providing effectively the functionality of a RNG.

There a two fundamental categories of RBGs, namely deterministic and non-deterministic. Using so-called seeds, deterministic random bit generators (DRBGs) use cryptographically secure algorithms to generate a sequence of random bits. This construction can also be referenced as pseudo random bit generator (PRNG).
On the other hand there are non-deterministic random bit generators (NRBGs), also called TRNGs.
The base for a NRBG is an unpredictable physical source that is not controllable by any human. Such an unpredictable source is commonly known as an entropy source. \cite[p.~62]{80090B:2018}

An example for a PRNG is deployed within the Java Runtime Environment, namely java.random. This PRNG will deliver the same sequence of numbers given the same seed on each call. \cite{kutschera:2019} 
A single photon source driven TRNG that utilizes quantum mechanics can be bought from ID Quantique \cite{idquantique:2016}. This specific type of TRNG is due to its quantum mechanics features also called QRNG, which we had the chance to utilize in our research \cite{kutschera:2019}. 

\subsection{Relevant Existing Random Number Generators}\label{sec:existing_RNGS}
In the following we list the most relevant RNGs which use entropy source in combination with modulo operations to generate random bits.

\subsubsection{Proposed RNG by Zhang}
Placing a human finger on the camera of a smartphone to create a TRNG was suggested by Zhang \cite{zhang:2014}. Utilizing the flashlight, the finger was illuminated in an unpredictable way. Applying the modulo two operation on the resulting data stream, the authors claim to have extracted 'entropy enhanced true random bits'. It is argued that the Bayer Mosaic Pattern, that describes the placement of the red, blue and green color sensors, with the postprocessing demosaicing process is harmful to randomness. \cite{zhang:2014}

\subsubsection{Proposed RNG by Leschiutta}
The motivation behind the RNG called 'BlueRand' was to send encrypted messages within arbitrary messaging services. The scheme behind BlueRand required the creation of a one-time pad. This was achieved by using the difference in the blue color channels of two independent pictures. The creator of BlueRand used ENT and RaBiGeTe to analyze the results. \cite{leschiutta:2015, leschiutta:2016}

\subsubsection{Proposed RNG by Chen}
This RNG fetched the coordinates (x,y) of video input. Then it set a threshold on which future values on a specific coordinate were dropped. Lastly, in correlation to the sampled audio the RNG applied a bitwise operation on the base rgb-color channels.
The author stated that the RNG passed all tests provided in the NIST Statistical Test Suite. \cite{chen:2013}

\subsubsection{Proposed RNG by Krhovj\'{a}k}
This specific RNG was based on the average entropy within testobjects \cite{krhovjak:2007}. The authors used Nokia N73, E-Ten X500 and E-Ten M700 as test objects. Four different methods were described: Processing raw values only, least significant color bit, color combination using XOR and Flip-Flop bit extraction. The authors argued that the most robust method was the Flip-Flop bit extraction where the modulo operation, $\pmod{2}$, was executed on each pixel. Eventually, the output of this method did not pass all tests of the NIST Statistical Test Suite.

\section{Implementation}\label{chap:implementation}
\subsection{ERBG Proof of Concept}\label{sec:erbg_implementation}
As within the proposed algorithm in Section~\ref{sec:proposed_algorithm}, can be seen that the main operation that is executed on every byte is $\pmod{2}$. A similar technique can be found in other RNGs \cite{leschiutta:2015,leschiutta:2016,zhang:2014,krhovjak:2007} but focused on the color channel. In this implementation of a proof of concept; however, on each read byte, a $\pmod{2}$ operation took place.  As it turned out that the first approximately 3250 bits represented the same zero values, it led to the decision to cut the first 4000 bits, which includes an arbitrary set margin of 23\%.

\subsubsection{Development}\label{subsec:development}
\begin{figure*}
    \centering
    \subfigure[Main View]{\label{fig:erbg_main_screen}\includegraphics[width=0.275\textwidth]{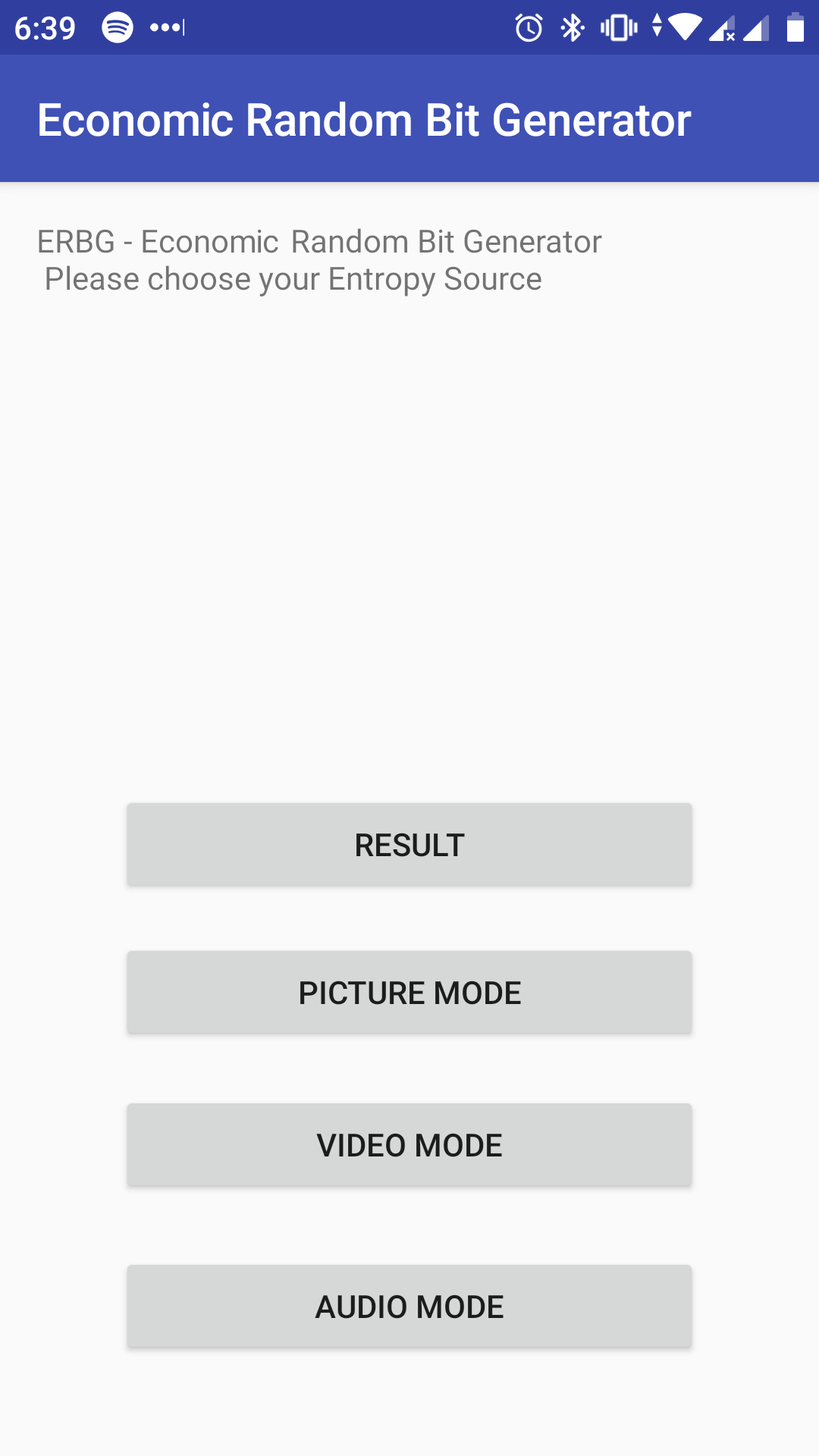}}
    \subfigure[Video View]{\label{fig:video_main_screen}\includegraphics[width=0.275\textwidth]{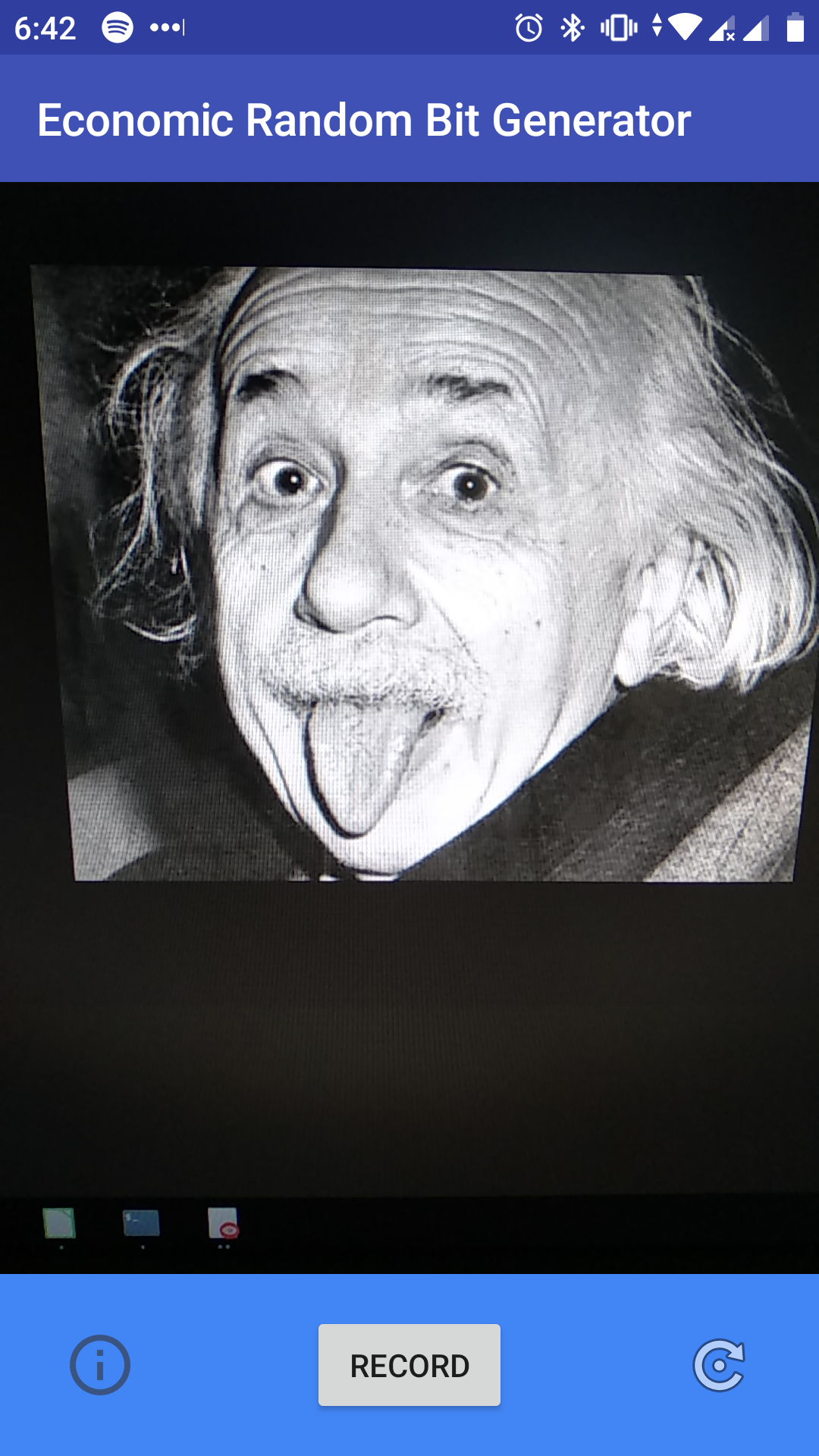}}
    \subfigure[Audio View]{\label{fig:audio_main_screen}\includegraphics[width=0.275\textwidth]{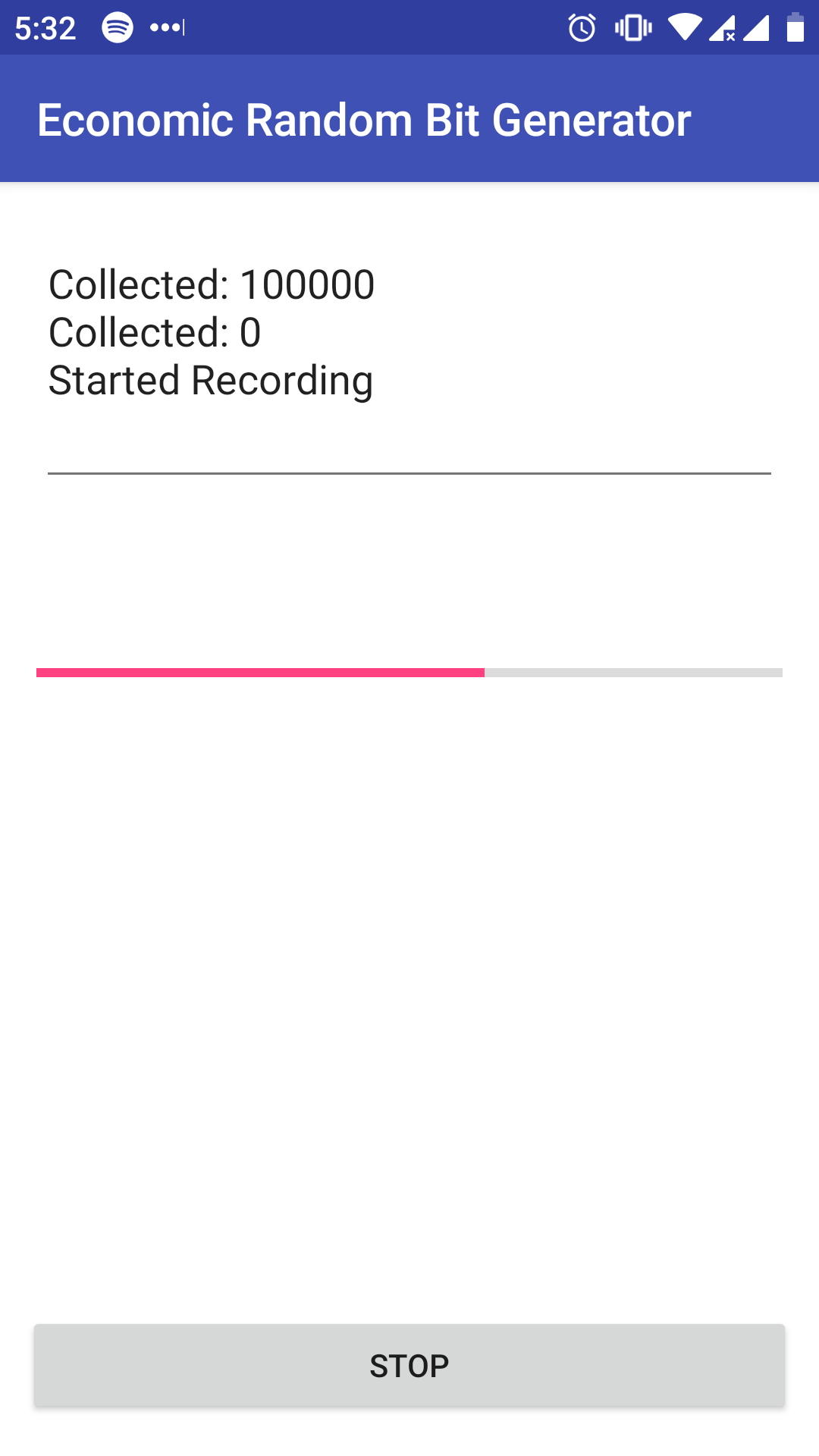}}
    \caption{A screenshot of the main, video, and audio view within the ERBG Android application, including graphically fixed ERBG naming error.}
\end{figure*}

The implementation of the ERBG prototype used the Model-View-Control pattern. \emph{MainActivity} is called during the start of the application and buttons for collection of the different entropy sources are initialized. It is worth mentioning that the \emph{Picture Mode}, as can be seen in Figure~\ref{fig:erbg_main_screen}, is not implemented but intended as a placeholder for further research on muons. The implementation using so-called Fragments allows users to go back and forth within the application without losing data. After an entropy source fetched raw data, the class \emph{RandomService} was called to execute the proposed algorithm within Section~\ref{sec:proposed_algorithm} on the fetched raw data. Eventually, the \emph{ResultFragment} showed parts of the collected data as well as very basic statistical information such as the ratio between zero and ones as well as the longest run on ones and zeros. In order to allow failure on execution, it was decided to write the output file in an incremental matter constantly onto the file system.    

\subsection{Proposed Algorithm}\label{sec:proposed_algorithm}
With respect to other research using the modulo operation within their dedicated own algorithms on color channels, we implemented within our prototype and experiment a rather simple but effective approach we want to propose below.\par

For each 8-bit byte block $b$ within an array of $n$ bytes data stream, after the byte block threshold count $tc$ is reached within that specific data stream array of $n$ bytes, the following operation $E$ is executed for $b$. Where $m$ is the position of created random bit sequence.    

\begin{equation}\label{eq:erbg_algo}
E_{m}(b_n) = \pmod{2} \Bigg(|b_{n}|\Bigg)
\end{equation}

\begin{figure}[h]
    \centering
    \includegraphics[width=0.7\columnwidth]{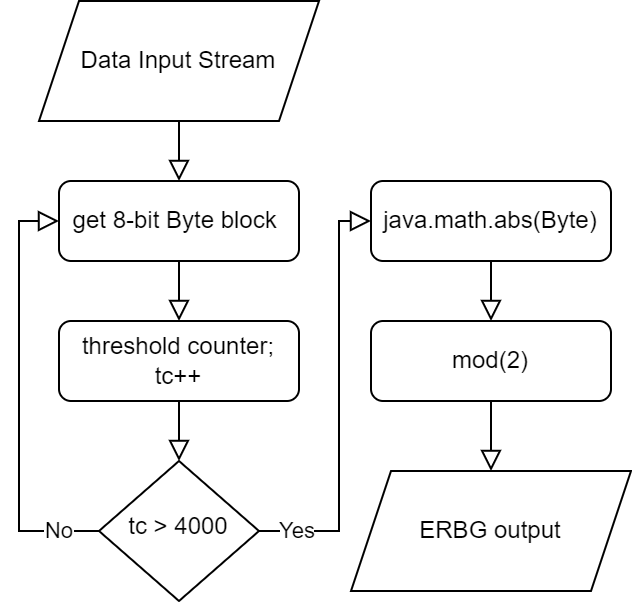}
    \caption{Flowchart of used algorithm within the ERBG.}
    \label{fig:erbg_flowchart}
\end{figure}

This can also be depicted as a flowchart as shown within Figure~\ref{fig:erbg_flowchart}.

\subsection{Test Parameters}\label{SEC_sts_params}
During our research, we found it often difficult to get an inside into the parameters used within the NIST Statistical Test Suite. In one occurrence, we were able to find a major execution error that resulted in crashes of the NIST Statistical Test Suite.\cite{kutschera:2019} For transparency and reproducibility following Table~\ref{tab:sts_test_param} states our used test parameters. 

\begin{table}[h]
    \centering
    \caption{Used parameters and settings to run the NIST Statistical Test Suite.} 
    \begin{tabular}{ l | l }
        \hline
        \rowcolor{gray5}\textbf{Parameter}    & \textbf{Value} \\
        \hline
        Length of a bitstream & $1 x 10^6$\\
        Amount of bitstreams & 100 \\
        Applied statistical tests & 1 - 15\\
        Input file format & ASCII \\
        \hline
        Block Frequency Test - block length(M)         & 128   \\
        NonOverlapping Template Test - block length(m) & 9     \\
        Overlapping Template Test - block length(m)    & 9     \\
        Approximate Entropy Test - block length(m)     & 10    \\
        Serial Test - block length(m)                  & 16    \\
        Linear Complexity Test - block length(M)       & 500   \\
        \hline
    \end{tabular}
    \label{tab:sts_test_param}
\end{table}

\section{Evaluation \& Results}\label{ch:evaluation}

\subsection{Test Setup, Environment \& Methods}\label{SEC_testsetup_and_environment}
During our experiment, three test devices consisting out of two models are used. The Xiaomi Mi A1 as well as the LG Nexus 5X both run on stock Android which basically means the operating system Android was not modified by the manufacturer. More hardware specifications can be found within Table~\ref{tbl:device_comp}. 

\begin{table}[H]
    \caption{Comparison of the specifications from the different used devices, LG Nexus 5X as well as Xiaomi Mi A1.\cite{devspec:2019}}
    \centering
    \begin{tabular}{l|rrrr}
        \hline
        \rowcolor{gray5}    Type/Detail         & Xiaomi Mi A1 & LG Nexus 5X \\
        \hline
        \hline
        \multicolumn{3}{c}{Primary Camera / Back Camera} \\
        \hline
        Image Sensor Type                       & PureCel                & CMOS BSI \\
        Image Sensor Manufacturer               & OmniVision             & Sony     \\
        Image Sensor Model                      & OV12A10                & IMX377   \\
        \hline
        \hline
        \multicolumn{3}{c}{Secondary Camera / Front Camera} \\
        \hline
        Image Sensor Type                       & CMOS BSI                 & CMOS BSI 2\\
        Image Sensor Manufacturer               & Samsung                  & OmniVision\\
        Image Sensor Model                      & S5K5E8                   & OV5693\\
        \hline
    \end{tabular}
    \label{tbl:device_comp}
\end{table}

During the experiment and ongoing evaluation of results, we found that similar scenarios generated passing and failing random bit sequences when tested with the NIST Statistical Test Suite. This led to the construction of an apparatus from cardboard that was able to hold two devices at the same time and let the operator trigger both devices at the almost same time. The apparatus consisted of a hole on the back for each smartphone to give a clear view for the lens. The front holes were made to allow the operator with spread fingers to trigger the devices at the very same seconds. The limitations of such an apparatus are that it cannot guarantee to trigger each phone at the very same millisecond. Another limitation is the parallax caused by placing each phone sideways next to each other. Concluding, we assume this setup to be sufficient enough to draw conclusions from it as the benefits of the apparatus caused the devices to experience the same movements, acceleration, and almost the same perspective with respect to the mentioned parallax.  
\begin{figure}
    \centering
    \subfigure[Apparatus Front]{\label{fig:apparatus_front}\includegraphics[width=0.20\textwidth]{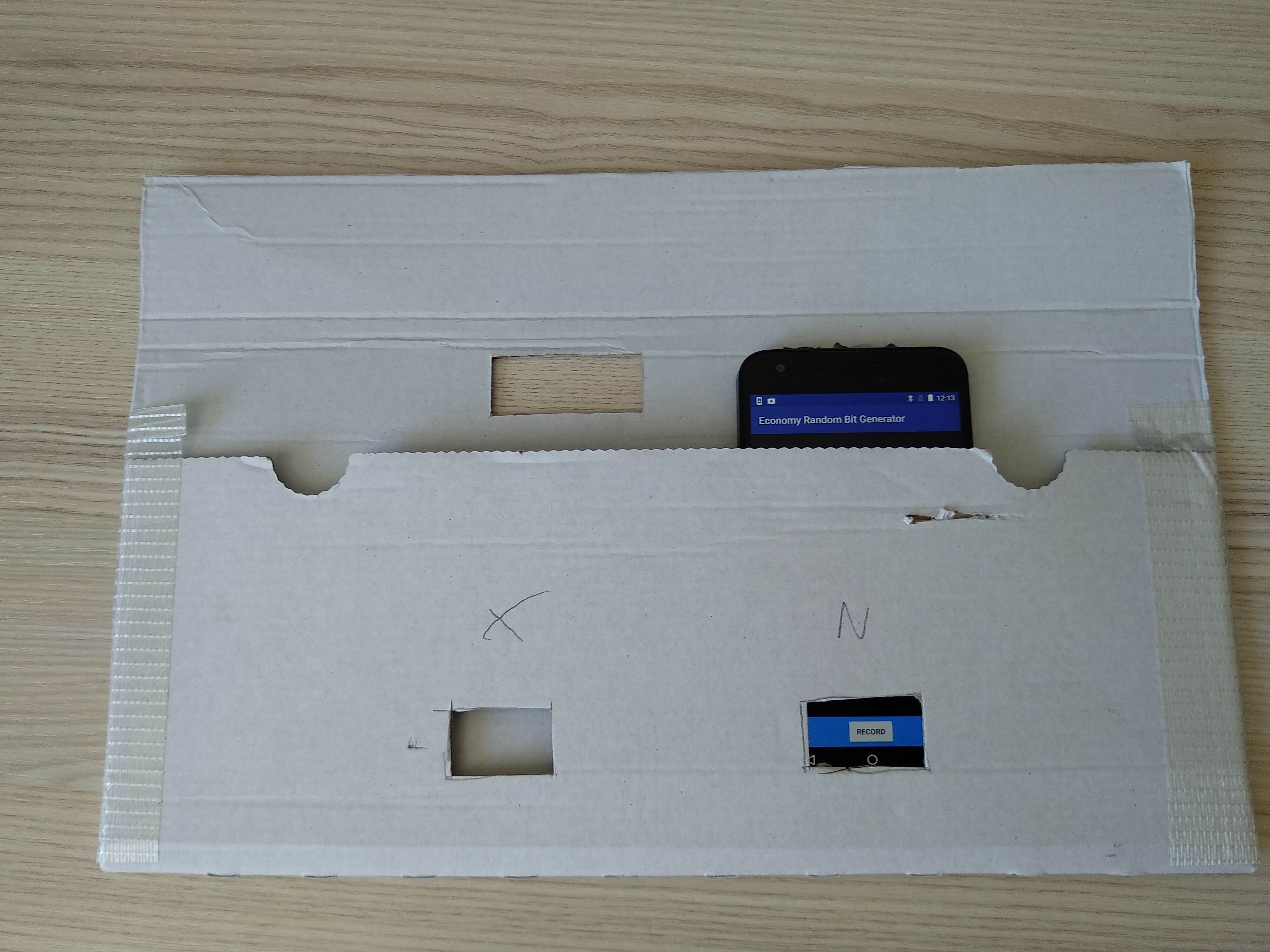}}
    \subfigure[Apparatus Back]{\label{fig:apparatus_back}\includegraphics[width=0.20\textwidth]{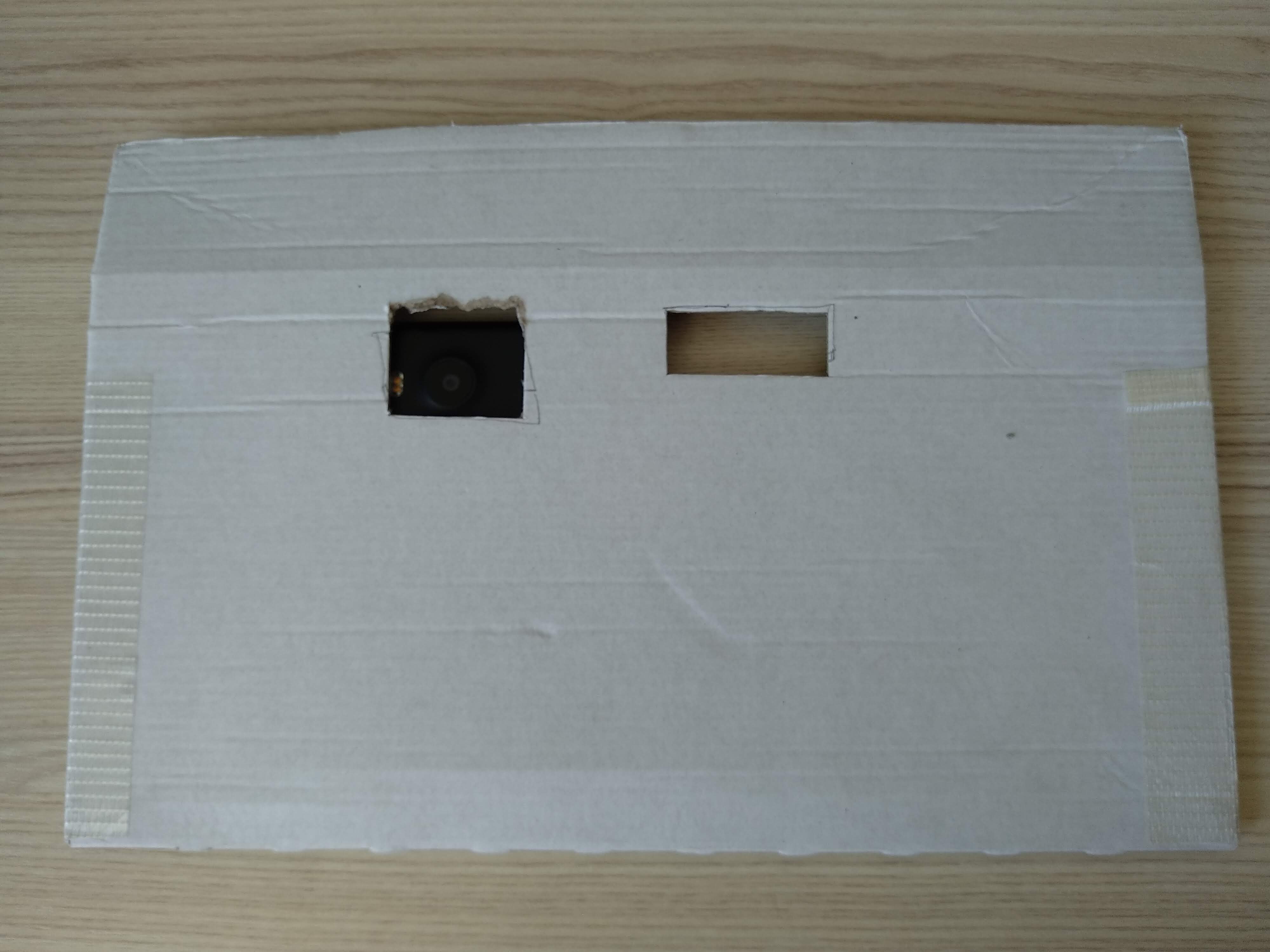}}
    \caption{Shows the apparatus used to execute the experiments.}
\end{figure}

\begin{figure*}
    \centering
    \subfigure[Fast Forest Circle Spin Test]{\label{fig:ffsc_1}\includegraphics[width=0.22\textwidth]{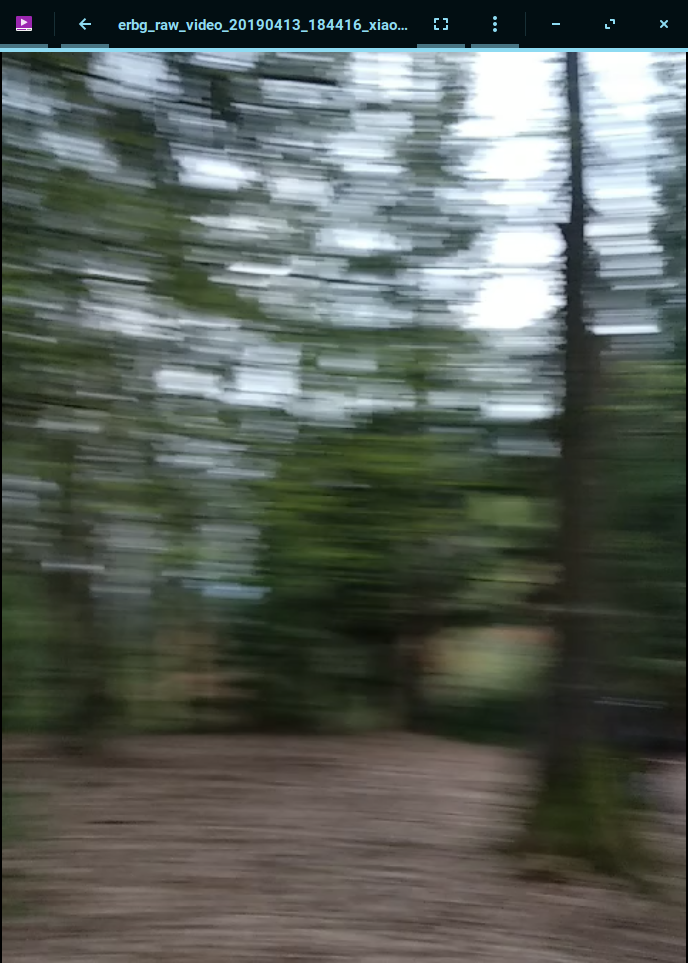}}
    \subfigure[Fast Forest Circle Spin Test]{\label{fig:ffsc_2}\includegraphics[width=0.22\textwidth]{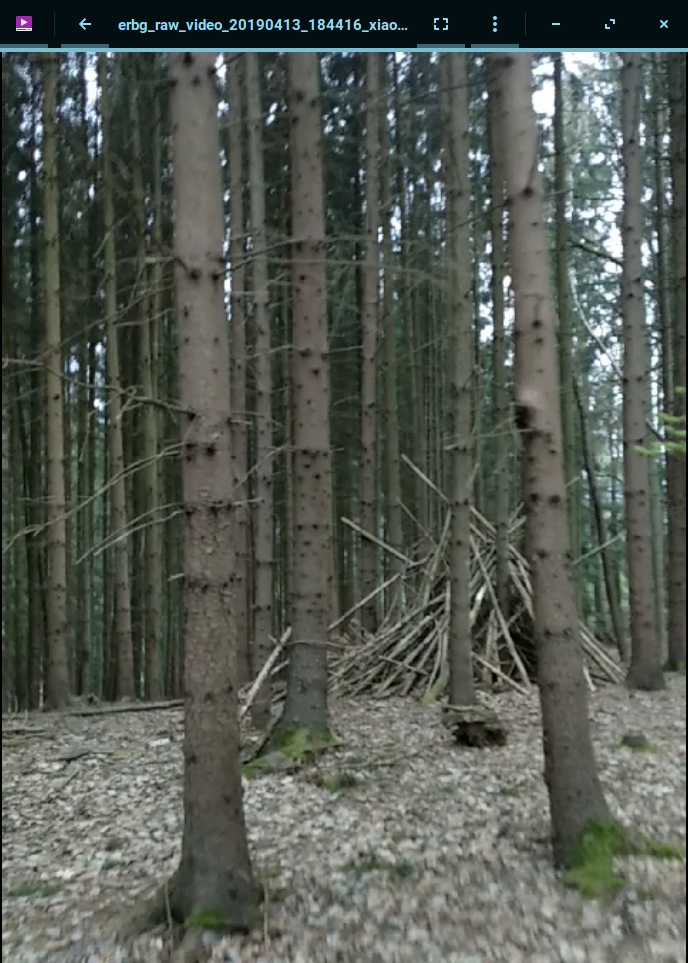}}
    \subfigure[Slow Meadow Linear Translation Test]{\label{fig:smlt_1}\includegraphics[width=0.22\textwidth]{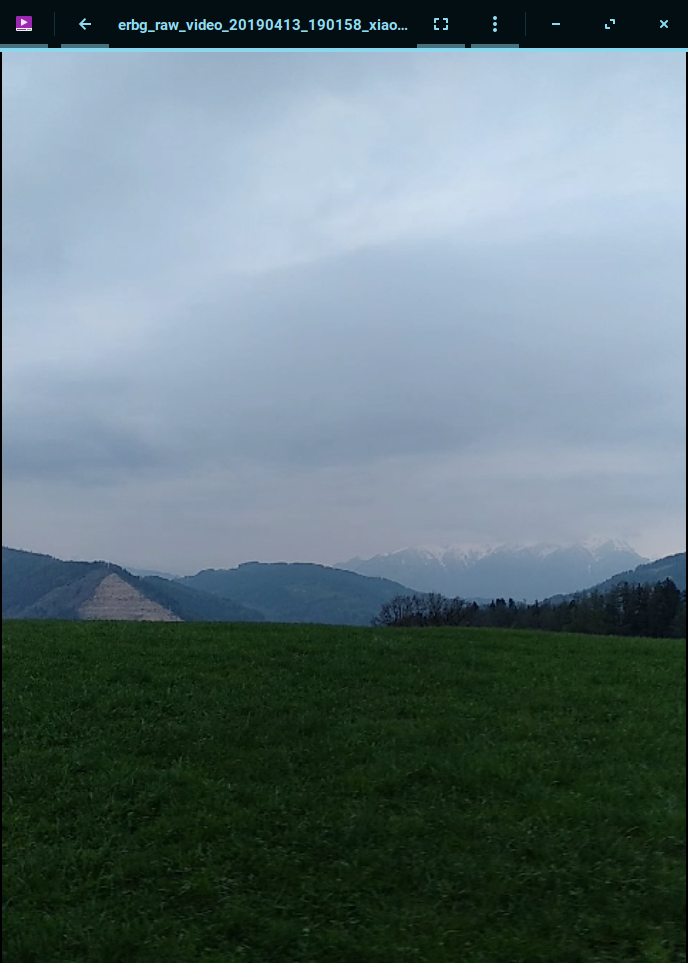}}
    \subfigure[Slow Meadow Linear Translation Test]{\label{fig:smlt_2}\includegraphics[width=0.22\textwidth]{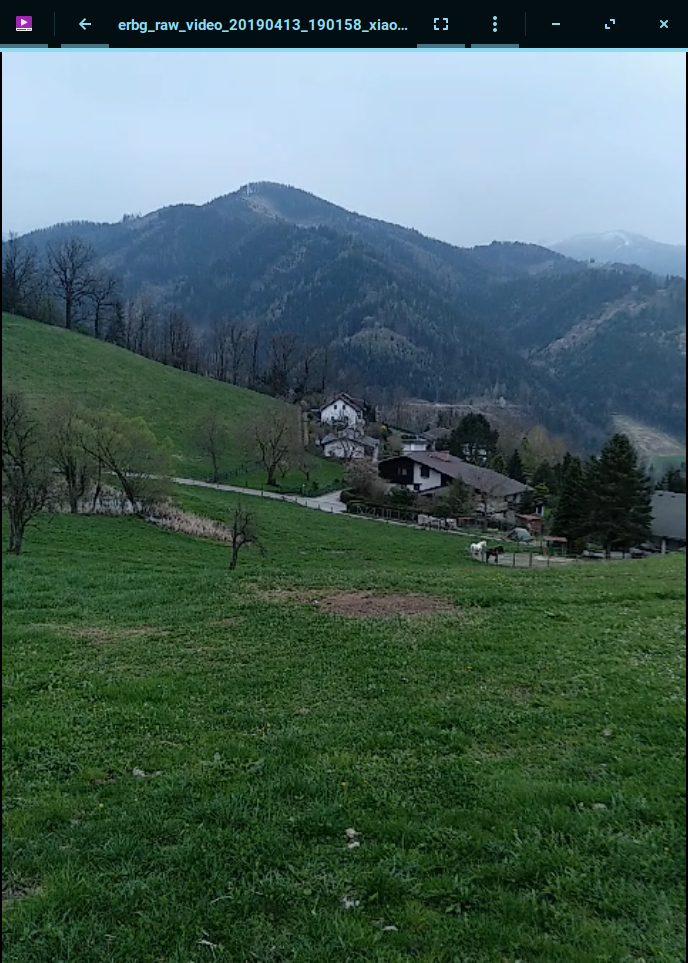}}
    \caption{Screenshot representing the two distinctive methods of collecting randomness using the ERBG.}
\end{figure*}

Under the usage of the aforementioned apparatus, we found two distinctive methods of how randomness can be sampled from the surroundings utilizing our ERBG prototype. First, during the collection process, a person has a lot of movement. For instance, one is spinning around the own axis within a forest. We call such collection method \emph{Fast Forest Circle Spin} (FFCS). Secondly, if someone does not move at all or at a very slow pace in almost the same direction, for example, walks in a straight line, thus translate, on a meadow, we call this method \emph{Slow Meadow Linear Translation} (SMLT). A sample of the FFCS method can be seen within Figures~\ref{fig:ffsc_1}~-~\ref{fig:ffsc_2}, whereas the SMLT method is visualized within Figures~\ref{fig:smlt_1}~-~\ref{fig:smlt_2}.

\subsection{Results}\label{sec:results}
Results have shown that random bit sequences extracted and produced by our ERBG prototype using the FFCS include a humongous amount of motion blur and pass 15 out of 15 test provided within the NIST Statistical Test Suite as can be seen within Table~\ref{tbl:xiaSte_FFCS}. Hence, the generated sequence of random bits can be considered to have good random properties. In contrast, if SMLT is used during collection, the tests are failing as can be seen within Tables~\ref{tbl:xiaSte_smlt}~-~\ref{tbl:nexSte_smlt}. Notice that failed tests are marked by the NIST Statistical Testsuite with an asterisk '*'.

\subsubsection{Extracted Randomness from Audio}
3 hours and 32 minutes needed to pass in order to capture a sequence of $10^{8}$ random bits from the microphone input. Moreover, one test did not pass the NIST Statistical Test Suite. The results were consistent throughout different devices. Excluding the downside of long duration and a single failed test, this method seems promising but needs further research. Eventually, we did not pursue this method further within our work. 

\subsubsection{ERBG - Fast Forest Circle Spin}\label{subsec:fast_forest_circle_spin}
The collection processes with our ERBG prototype took 2:50 minutes, during which the operator rotated fast around the own axis. That actions gave the captured video high amount of motion blur, as can be seen in Figure~\ref{fig:ffsc_1}. It is worth mentioning that due to human limitations the operator cannot spin arbitrary long which results in momentary parts of the capturing process being without the fast movements, as Figure~\ref{fig:ffsc_2} clearly shows.
In the end, this specific test method passes all 15 tests of the NIST Statistical Test Suite. According to specifications within the NIST Statistical Test Suite at least 96 out of 100 bit-streams need to have a P-value between 0.01 to 0.99 to be considered unanimous but random enough. 
With respect to the pass rate, an exception of the random excursion and random excursion variant exists as at least 59 out of 62 bit-streams need to pass. 

\begin{table}[h]
    \caption{NIST Statistical Test Suite results for random bit sequence with our ERBG prototype on the Xiaomi Mi A1 using the \textbf{``Fast~Forest~Circle~Spin''} method.}
    \centering
    \begin{tabular}{lcl|l}
        \hline
        \rowcolor{gray5}P-VALUE & PROPORTION & STATISTICAL TEST & PASS  \\
		\hline
		\cellcolor{testpassed} 0.319084 & \cellcolor{testpassed} 97/100  & Frequency               & \cellcolor{testpassed} YES \\
		\cellcolor{testpassed} 0.514124 & \cellcolor{testpassed} 96/100  & BlockFrequency          & \cellcolor{testpassed} YES \\
		\cellcolor{testpassed} 0.153763 & \cellcolor{testpassed} 96/100  & CumulativeSums          & \cellcolor{testpassed} YES \\
		\cellcolor{testpassed} 0.401199 & \cellcolor{testpassed} 97/100  & CumulativeSums          & \cellcolor{testpassed} YES \\
		\cellcolor{testpassed} 0.040108 & \cellcolor{testpassed} 99/100  & Runs                    & \cellcolor{testpassed} YES \\
		\cellcolor{testpassed} 0.115387 & \cellcolor{testpassed} 99/100  & LongestRun              & \cellcolor{testpassed} YES \\
		\cellcolor{testpassed} 0.494392 & \cellcolor{testpassed} 100/100 & Rank                    & \cellcolor{testpassed} YES \\
		\cellcolor{testpassed} 0.455937 & \cellcolor{testpassed} 100/100 & FFT                     & \cellcolor{testpassed} YES \\
		\cellcolor{testpassed} 0.035174 & \cellcolor{testpassed} 99/100  & NonOverlappingTemplate  & \cellcolor{testpassed} YES \\
		\cellcolor{testpassed} 0.040108 & \cellcolor{testpassed} 99/100  & OverlappingTemplate     & \cellcolor{testpassed} YES \\
		\cellcolor{testpassed} 0.366918 & \cellcolor{testpassed} 99/100  & Universal               & \cellcolor{testpassed} YES \\
		\cellcolor{testpassed} 0.181557 & \cellcolor{testpassed} 99/100  & ApproximateEntropy      & \cellcolor{testpassed} YES \\
		\cellcolor{testpassed} 0.025193 & \cellcolor{testpassed} 59/59   & RandomExcursions        & \cellcolor{testpassed} YES \\
		\cellcolor{testpassed} 0.071177 & \cellcolor{testpassed} 58/59   & RandomExcursionsVariant & \cellcolor{testpassed} YES \\
		\cellcolor{testpassed} 0.455937 & \cellcolor{testpassed} 100/100 & Serial                  & \cellcolor{testpassed} YES \\
		\cellcolor{testpassed} 0.534146 & \cellcolor{testpassed} 98/100  & Serial                  & \cellcolor{testpassed} YES \\
		\cellcolor{testpassed} 0.137282 & \cellcolor{testpassed} 99/100  & LinearComplexity        & \cellcolor{testpassed} YES \\ \hline
    \end{tabular}
    \label{tbl:xiaSte_FFCS}
\end{table}

\subsubsection{ERBG - Slow Meadow Linear Translation}\label{subsec:slow_meadow_linear_translation}
Collected data using our ERBG prototype and the SMLT method show several failures within the results from the NIST Statistical Test Suite. As for all experiment runs the aforementioned apparatus was also used in this experiment run. Eventually, the results as seen within Table~\ref{tbl:xiaSte_smlt}~-~\ref{tbl:nexSte_smlt} were collected at the same time. After analyzing the captured video stream, the view seems steady and the color seldom changes. As for the strong negative results from the NIST Statistical Test Suite, and the clear indication in relation to color/picture/scenery changes we decided against a comparison of the color histogram. Interestingly, the random sequence collected with LG Nexus 5X as shown in Table~\ref{tbl:nexSte_smlt} fail 73,33\% of all tests whereas the random sequence collected with Xiaomi Mi A1 fails in only 33,33\%.

\begin{table}[H]
\caption{NIST Statistical Test Suite results for random bit sequence with our ERBG prototype on the Xiaomi Mi A1 using the \textbf{``Slow~Meadow~Linear~Translation''} method.}
\centering
    \begin{tabular}{lcl|l}
        \hline
        \rowcolor{gray5}P-VALUE  & PROPORTION & STATISTICAL TEST   & PASS  \\
        \hline
        \cellcolor{testfailed} 0*        & \cellcolor{testpassed} 97/100     & Frequency               & \cellcolor{testfailed} NO   \\
        \cellcolor{testfailed} 0.000009* & \cellcolor{testpassed} 97/100     & BlockFrequency          & \cellcolor{testfailed} NO   \\
        \cellcolor{testfailed} 0*        & \cellcolor{testpassed} 96/100     & CumulativeSums          & \cellcolor{testfailed} NO   \\
        \cellcolor{testfailed} 0*        & \cellcolor{testpassed} 96/100     & CumulativeSums          & \cellcolor{testfailed} NO   \\
        \cellcolor{testpassed} 0.55442  & \cellcolor{testpassed} 100/100    & Runs                    & \cellcolor{testpassed} YES  \\
        \cellcolor{testpassed} 0.719747 & \cellcolor{testpassed} 99/100     & LongestRun              & \cellcolor{testpassed} YES  \\
        \cellcolor{testpassed} 0.145326 & \cellcolor{testpassed} 98/100     & Rank                    & \cellcolor{testpassed} YES  \\
        \cellcolor{testpassed} 0.137282 & \cellcolor{testpassed} 100/100    & FFT                     & \cellcolor{testpassed} YES  \\
        \cellcolor{testpassed} 0.191687 & \cellcolor{testfailed} 95/100*     & NonOverlappingTemplate  & \cellcolor{testfailed} NO   \\
        \cellcolor{testpassed} 0.883171 & \cellcolor{testpassed} 98/100     & OverlappingTemplate     & \cellcolor{testpassed} YES  \\
        \cellcolor{testpassed} 0.955835 & \cellcolor{testpassed} 99/100     & Universal               & \cellcolor{testpassed} YES  \\
        \cellcolor{testpassed} 0.366918 & \cellcolor{testpassed} 97/100     & ApproximateEntropy      & \cellcolor{testpassed} YES  \\
        \cellcolor{testpassed} 0.227773 & \cellcolor{testpassed} 42/43      & RandomExcursions        & \cellcolor{testpassed} YES  \\
        \cellcolor{testpassed} 0.113706 & \cellcolor{testpassed} 42/43      & RandomExcursionsVariant & \cellcolor{testpassed} YES  \\
        \cellcolor{testpassed} 0.051942 & \cellcolor{testpassed} 99/100     & Serial                  & \cellcolor{testpassed} YES  \\
        \cellcolor{testpassed} 0.455937 & \cellcolor{testpassed} 100/100    & Serial                  & \cellcolor{testpassed} YES  \\
        \cellcolor{testpassed} 0.12962  & \cellcolor{testpassed} 99/100     & LinearComplexity        & \cellcolor{testpassed} YES  \\
        \hline
    \end{tabular}
\label{tbl:xiaSte_smlt}
\end{table}

\begin{table}[ht]
    \caption{NIST Statistical Test Suite results for random bit sequence with our ERBG prototype on the LG Nexus 5X using the \textbf{``Slow~Meadow~Linear~Translation''} method.}
    \centering
    \begin{tabular}{lcl|l}
        \hline
        \rowcolor{gray5}P-VALUE  & PROPORTION & STATISTICAL TEST   & PASS  \\
        \hline
        \cellcolor{testfailed} 0*        & \cellcolor{testfailed} 3/100*      & Frequency               & \cellcolor{testfailed} NO   \\
        \cellcolor{testfailed} 0*        & \cellcolor{testfailed} 0/100*      & BlockFrequency          & \cellcolor{testfailed} NO   \\
        \cellcolor{testfailed} 0*        & \cellcolor{testfailed} 0/100*      & CumulativeSums          & \cellcolor{testfailed} NO   \\
        \cellcolor{testfailed} 0*        & \cellcolor{testfailed} 0/100*      & CumulativeSums          & \cellcolor{testfailed} NO   \\
        \cellcolor{testfailed} 0*        & \cellcolor{testfailed} 1/100*      & Runs                    & \cellcolor{testfailed} NO   \\
        \cellcolor{testfailed} 0*        & \cellcolor{testfailed} 67/100*     & LongestRun              & \cellcolor{testfailed} NO   \\
        \cellcolor{testpassed} 0.289667 & \cellcolor{testpassed} 100/100    & Rank                    & \cellcolor{testpassed} YES  \\
        \cellcolor{testpassed} 0.867692 & \cellcolor{testpassed} 97/100     & FFT                     & \cellcolor{testpassed} YES  \\
        \cellcolor{testfailed} 0*        & \cellcolor{testfailed} 2/100*      & NonOverlappingTemplate  & \cellcolor{testfailed} NO   \\
        \cellcolor{testfailed} 0*        & \cellcolor{testfailed} 20/100*     & OverlappingTemplate     & \cellcolor{testfailed} NO   \\
        \cellcolor{testfailed} 0*        & \cellcolor{testfailed} 57/100*     & Universal               & \cellcolor{testfailed} NO   \\
        \cellcolor{testfailed} 0*        & \cellcolor{testfailed} 17/100*     & ApproximateEntropy      & \cellcolor{testfailed} NO   \\
        \cellcolor{testfailed} ----     & \cellcolor{testfailed} ------     & RandomExcursions        & \cellcolor{testfailed} NO   \\
        \cellcolor{testfailed} ----     & \cellcolor{testfailed} ------     & RandomExcursionsVariant & \cellcolor{testfailed} NO   \\
        \cellcolor{testfailed} 0*        & \cellcolor{testfailed} 55/100*     & Serial                  & \cellcolor{testfailed} NO   \\
        \cellcolor{testpassed} 0.657933 & \cellcolor{testpassed} 96/100     & Serial                  & \cellcolor{testpassed} YES  \\
        \cellcolor{testpassed} 0.171867 & \cellcolor{testpassed} 100/100    & LinearComplexity        & \cellcolor{testpassed} YES  \\
        \hline
    \end{tabular}
    \label{tbl:nexSte_smlt}
\end{table}

\section{Conclusion and Outlook}\label{chap:conclusion}
Randomness fascinates human beings at least since the Middle Ages \cite{batanero:1998}.
Nowadays, we draw not the future from randomness but rather protect our future using randomness. As certain cryptographic schemes try to protect crucial private and/or business information from parties not eligible in receiving certain information. Moreover, randomness is defined as the lack of patterns and predictability. With that definition and a deeper understanding of patterns, it became more and more difficult in creating a strong RNG.\par

We conclude that, given the complex topic with our yet so simple approach, our proposed ERBG indicates sufficient strong randomness when tested against the NIST Statistical Test Suite. However, our work also has its limitations as incorrect usage can undermine the quality of collected random bit sequences significantly. We also found that certain devices had weak random bit sequences and were not linked to any video configuration like frame-rate, codec, codec-profile, resolution, stabilization provided by the operating system nor the image-sensor type. Such behavior did not occur with the, not further pursued, audio collection of randomness. 

We have shown a clear indication that the ERBG prototype has potential for further research. We also think it is worth revisiting the cosmic ray experiment, which failed due to organizational errors during execution \cite{kutschera:2019}. With more focused knowledge, existing successful research projects\cite{whiteson:2015, homola:2020, deco:2015}, the implementation of an cosmic ray, muon powered entropy source, and feedback from this work, research on the ERBG shall be continued.

\bibliographystyle{IEEEtran}
\bibliography{IEEEabrv,references}

\begin{thebibliography}{10}
\providecommand{\url}[1]{#1}
\csname url@samestyle\endcsname
\providecommand{\newblock}{\relax}
\providecommand{\bibinfo}[2]{#2}
\providecommand{\BIBentrySTDinterwordspacing}{\spaceskip=0pt\relax}
\providecommand{\BIBentryALTinterwordstretchfactor}{4}
\providecommand{\BIBentryALTinterwordspacing}{\spaceskip=\fontdimen2\font plus
\BIBentryALTinterwordstretchfactor\fontdimen3\font minus
  \fontdimen4\font\relax}
\providecommand{\BIBforeignlanguage}[2]{{%
\expandafter\ifx\csname l@#1\endcsname\relax
\typeout{** WARNING: IEEEtran.bst: No hyphenation pattern has been}%
\typeout{** loaded for the language `#1'. Using the pattern for}%
\typeout{** the default language instead.}%
\else
\language=\csname l@#1\endcsname
\fi
#2}}
\providecommand{\BIBdecl}{\relax}
\BIBdecl

\bibitem{idquantique:2016}
{ID Quantique, Swiss Quantum}, \emph{Quantis, When Random Numbers Cannot be
  left to chance}.\hskip 1em plus 0.5em minus 0.4em\relax ID Quantique SA,
  Chemin de la Marbrerie 3, Geneva, Switzerland, 06 2016.

\bibitem{80057rev5:2020}
E.~Barker, \emph{Recommendation for Key Managment, NIST Special Publication
  800-57 Part1 Revision 5}.\hskip 1em plus 0.5em minus 0.4em\relax 100 Bureau
  Drive, Gaithersburg, Maryland 20899, Unites States of America: National
  Institute of Technology, Computer Security Division, May 2020.

\bibitem{kutschera:2019}
S.~Kutschera, \emph{Construction and Evaluation of a Non-Deterministic Random
  Bit Generator with Mobile Sensors as Entropy Source}.\hskip 1em plus 0.5em
  minus 0.4em\relax FH JOANNEUM GmbH, Kapfenberg, 2019.

\bibitem{80022:2010}
A.~Rukhin, R.~Soto, J.~Nechvatal, M.~Smid, E.~Barker, S.~Leigh, M.~Levenson,
  M.~Vangel, D.~Banks, A.~Heckert, J.~Dray, and S.~Vo, \emph{A Statistical Test
  Suite for Random and Pseudorandom Number Generators for Cryptographic
  Applications, NIST Special Publication 800-22 Revision 1}.\hskip 1em plus
  0.5em minus 0.4em\relax 100 Bureau Drive, Gaithersburg, Maryland 20899,
  Unites States of America: National Institute of Technology, Computer Security
  Division and Statistical Engineering Divison, April 2010.

\bibitem{batanero:1998}
C.~Batanero, D.~R. Green, and L.~R. Serrano, ``Randomness, its meanings and
  educational implications,'' \emph{International Journal of Mathematical
  Education in Science and Technology}, vol.~29, no.~1, pp. 113--123, 1998.

\bibitem{bennet:1993}
D.~Bennet, \emph{The development of the mathematical concept of
  randomness}.\hskip 1em plus 0.5em minus 0.4em\relax New York Univeristy,
  1993.

\bibitem{poincare:1936}
H.~Poincare, \emph{Foundation of Science:Chance Translated by J.R.
  Newman}.\hskip 1em plus 0.5em minus 0.4em\relax The World of Mathematics,
  1936.

\bibitem{80090B:2018}
M.~S. Turan, E.~Barker, J.~Kelsey, K.~A. McKay, M.~L. Baish, and M.~Boyle,
  \emph{Recommendation for the Entropy Sources Used for Random Bit Generation,
  NIST Special Publication 800-90B}.\hskip 1em plus 0.5em minus 0.4em\relax 100
  Bureau Drive, Gaithersburg, Maryland 20899, Unites States of America:
  National Institute of Technology, Computer Security Division, January 2018.

\bibitem{zhang:2014}
\BIBentryALTinterwordspacing
X.~Zhang, L.~Qi, Z.~Tang, and Y.~Zhang, ``Portable true random number generator
  for personal encryption application based on smartphone camera,''
  \emph{Electronics Letters}, vol.~50, no.~24, pp. 1841--1843, 2014. [Online].
  Available:
  \url{https://ietresearch.onlinelibrary.wiley.com/doi/abs/10.1049/el.2014.2870}
\BIBentrySTDinterwordspacing

\bibitem{leschiutta:2015}
R.~Leschiutta, \emph{PARAPPNOID: Un’Applicazione di Messaggisstica Basata au
  One-Time Pad}.\hskip 1em plus 0.5em minus 0.4em\relax Universita Degli Studi
  di Udine, Dipartimento di Scienze Matematiche, Informatiche e Fisiche, 2015.

\bibitem{leschiutta:2016}
\BIBentryALTinterwordspacing
------, \emph{Java True Random Number Generator (TRNG) that uses JPEG images as
  entropy source.}\hskip 1em plus 0.5em minus 0.4em\relax prgpascal@Github.com,
  2016. [Online]. Available: \url{https://github.com/prgpascal/bluerand}
\BIBentrySTDinterwordspacing

\bibitem{chen:2013}
I.-T. Chen, ``Random numbers generated from audio and video sources,''
  \emph{Hindawi Publishing Corporation Mathematical Problems in Engineering},
  vol. Volume 2013, Article ID 285373, 7 pages, pp. 1827--1832, July 2013.

\bibitem{krhovjak:2007}
J.~Krhovj\'{a}k, P.~\v{S}venda, and V.~Maty\'{a}\v{s}, ``The sources of
  randomness in mobile devices.''\hskip 1em plus 0.5em minus 0.4em\relax
  Reykjavik University, October 2007, nordsec 2007: The 12th Nordic Conference
  on Secure IT Systems.

\bibitem{devspec:2019}
\BIBentryALTinterwordspacing
DeviceSpecifications.com, \emph{Comparison between LG Nexus 5X and Xiaomi Mi
  A1}.\hskip 1em plus 0.5em minus 0.4em\relax devicespecifications.com, 4 2019.
  [Online]. Available:
  \url{https://www.devicespecifications.com/en/comparison/bf0ed3892}
\BIBentrySTDinterwordspacing

\bibitem{whiteson:2015}
D.~Whiteson, M.~Mulhearn, C.~Shimmin, K.~Cranmer, K.~Brodie, and D.~Burns,
  \emph{Observing Ultra-High Energy Cosmic Rays with Smartphones}.\hskip 1em
  plus 0.5em minus 0.4em\relax arXiv, Cornell University, 10 2015.

\bibitem{homola:2020}
P.~Homola, D.~Beznosko, G.~Bhatta, {\L}.~Bibrzycki, M.~Borczy{\'{n}}ska,
  {\L}.~Bratek, N.~Budnev, D.~Burakowski, D.~E. Alvarez-Castillo, K.~A.
  Cheminant, A.~{\'{C}}wik{\l}a, P.~Dam-o, N.~Dhital, A.~R. Duffy,
  P.~G{\l}ownia, K.~Gorzkiewicz, D.~G{\'{o}}ra, A.~C. Gupta,
  Z.~Hl{\'{a}}vkov{\'{a}}, M.~Homola, J.~Ja{\l}ocha, R.~Kami{\'{n}}ski,
  M.~Karbowiak, M.~Kasztelan, R.~Kierepko, M.~Knap, P.~Kov{\'{a}}cs,
  S.~Kuli{\'{n}}ski, B.~{\L}ozowski, M.~Magry{\'{s}}, M.~V. Medvedev,
  J.~M{k{e}}drala, J.~W. Mietelski, J.~Miszczyk, A.~Mozgova, A.~Napolitano,
  V.~Nazari, Y.~J. Ng, M.~Nied{\'{z}}wiecki, C.~Oancea, B.~Ogan, G.~Opi{\l}a,
  K.~Oziomek, M.~Pawlik, M.~Piekarczyk, B.~Poncyljusz, J.~Pryga, M.~Rosas,
  K.~Rzecki, J.~Zamora-Saa, K.~Smelcerz, K.~Smolek, W.~Stanek, J.~Stasielak,
  S.~Stuglik, J.~Sulma, O.~Sushchov, M.~Svanidze, K.~M. Tam, A.~Tursunov, J.~M.
  Vaquero, T.~Wibig, and K.~W. Wo{\'{z}}niak, ``Cosmic-ray extremely
  distributed observatory,'' \emph{Symmetry}, vol.~12, no.~11, p. 1835, nov
  2020.

\bibitem{deco:2015}
\BIBentryALTinterwordspacing
J.~Vandenbroucke, S.~Bravo, P.~Karn, M.~Meehan, M.~Plewa, T.~Ruggles,
  D.~Schultz, J.~Peacock, and A.~L. Simons, ``Detecting particles with cell
  phones: the distributed electronic cosmic-ray observatory,'' 2015. [Online].
  Available: \url{https://arxiv.org/abs/1510.07665}
\BIBentrySTDinterwordspacing

\end{thebibliography}
\end{document}